\documentclass[reprint,aps,prd,superscriptaddress,preprintnumbers,nofootinbib]{revtex4-1}

\usepackage{amsfonts}
\usepackage{amsmath}
\usepackage{amsthm}
\usepackage{amssymb}
\usepackage{appendix}
\usepackage{array}
\usepackage{dcolumn}
\usepackage{placeins}
\usepackage{comment}
\usepackage[svgnames]{xcolor}
\usepackage[hidelinks]{hyperref}
\hypersetup{
    colorlinks=true,
    linkcolor=NavyBlue,
    filecolor=NavyBlue,
    urlcolor=NavyBlue,
    citecolor=NavyBlue,
    }

\usepackage{graphics}
\usepackage{tikz}
\usepackage{graphicx}
\graphicspath{{plots/}}

\usetikzlibrary{positioning}
\usepackage{caption}
\usepackage{subcaption}
\usepackage{adjustbox}
\usepackage{gensymb}
\usepackage{breqn}
\usepackage{braket}
\usepackage{enumitem}
\usepackage{makecell}
\usepackage{ragged2e}
\usepackage{comment}
\usepackage{placeins}

\def\Zprime{$Z^{\prime}$ }
\begin{document}

\title{$E_6$ Models in Light of Precision $M_W$ Measurements}
\author{Vernon Barger}
\email{barger@pheno.wisc.edu}
\affiliation{Department of Physics$,$ \\University of Wisconsin$,$ Madison$,$ WI 53706}
\author{Cash Hauptmann}
\email{chauptmann2@huskers.unl.edu}
\author{Peisi Huang}
\email{peisi.huang@unl.edu}
\affiliation{Department of Physics  and Astronomy$,$\\  University of Nebraska$,$ Lincoln$,$ NE 68588}
\author{Wai-Yee Keung}
\email{keung@uic.edu}
\affiliation{Physics Department$,$\\
University of Illinois at Chicago$,$ IL 60607}
\begin{abstract}
   
    We propose a solution to the recent $W$ mass measurement by embedding the Standard Model within $E_6$ models. The presence of a new $U(1)$ group shifts the $W$ boson mass at the tree level and introduces a new gauge boson \Zprime which has been searched for at collider experiments. In this article, we identify the parameter space that explains the new $W$ mass measurement and is consistent with current experimental \Zprime searches. As $U(1)$ extensions can be accommodated in supersymmetric models, we also consider the supersymmetric scenario of $E_6$ models, and show that a 125 GeV Higgs may be easily achieved in such settings.
   
\end{abstract}
\maketitle

\section{Introduction}

Precision measurements have been crucial in testing physics beyond the Standard Model~(SM). In recent years, tensions between theory and experiment have been building with the muon $g-2$ measurement~\cite{Bennett:2006fi,g-2}, flavor anomalies~\cite{LHCb:2013ghj,LHCb:2014vgu,LHCb:2015svh,LHCb:2017avl,Belle:2019oag,LHCb:2021trn}, and most recently the $W$ boson mass measurement by the CDF collaboration~\cite{CDF}. The CDF II experiment measured the $W$ boson mass to be
\begin{equation}
    M_{W}^{\text{CDF}} = 80.4335 \pm 0.0094 \text{ GeV},
\end{equation} 
which deviates from the SM prediction~\cite{PDG} by about $7\sigma$,

\begin{equation} \label{eq:CDF}
    \delta M_W \equiv M_W^{\text{CDF}} - M_W^{\text{SM}} \approx 76 \pm 11 \text{ MeV}.
\end{equation}
This measurement has increased the tension between the SM and previous Tevatron measurements~\cite{CDF:2012gpf,D0:2012kms}, but is also in tension with the previous world average by more than 2$\sigma$~\cite{PDG}. The tension between various experiments can be from unknown systematic uncertainties, which is beyond the scope of this study. 

In this article, we focus on the compelling possibility that the deviation of results between the new CDF experiment, along with previous Tevatron experiments, and the SM predictions is a hint of new physics beyond the SM~\cite{Cheung2022isodoublet,Lu:2022bgw,DiLuzio:2022xns,Song:2022xts,Sakurai:2022hwh,Cheng:2022jyi,Bahl:2022xzi,Heo:2022dey,Biekotter:2022abc,Du:2022brr,Han:2022juu,Ahn:2022xeq,FileviezPerez:2022lxp,Ghoshal:2022vzo,Kanemura:2022ahw,Popov:2022ldh,Arcadi:2022dmt,Ghorbani:2022vtv,Lee:2022gyf,Heeck:2022fvl,Abouabid:2022lpg,Benbrik:2022dja,Kim:2022hvh,Strumia:2022qkt,deBlas:2022hdk,Yang:2022gvz,Yuan:2022cpw,Athron:2022qpo,Fan:2022dck,Babu:2022pdn,Heckman:2022the,Gu:2022htv,Athron:2022isz,Asadi:2022xiy,Paul:2022dds,Bagnaschi:2022whn,Lee:2022nqz,Liu:2022jdq,Fan:2022yly,Balkin:2022glu,Endo:2022kiw,Crivellin:2022fdf,Han:2022juu,Blennow:2022yfm,Cacciapaglia:2022xih,Tang:2022pxh,Zhu:2022tpr,Zheng:2022irz,Krasnikov:2022xsi,Arias-Aragon:2022ats,Du:2022pbp,Kawamura:2022uft,Nagao:2022oin,Zhang:2022nnh,Carpenter:2022oyg,Chowdhury:2022moc,Borah:2022obi,Zeng:2022lkk,Du:2022fqv,Bhaskar:2022vgk,Baek:2022agi,Cao:2022mif,Borah:2022zim,Batra:2022org,Almeida:2022lcs,Cheng:2022aau,Batra:2022pej,Benbrik:2022dja,Cai:2022cti,Zhou:2022cql,Gupta:2022lrt,Wang:2022dte,Barman:2022qix,Kim:2022xuo,Dcruz:2022dao,Isaacson:2022rts,Chowdhury:2022dps,Kim:2022zhj,Gao:2022wxk,Lazarides:2022spe,Rizzo:2022jti,VanLoi:2022eir,YaserAyazi:2022tbn,Chakrabarty:2022voz,CentellesChulia:2022vpz,Nagao:2022dgl,Bahl:2022gqg,Arora:2022uof,Heinemeyer:2022ith,Benakli,https://doi.org/10.48550/arxiv.2207.01465}. In particular, we focus on a possible tree-level modification to the $W$ boson mass coming from an extension the SM gauge group. The simplest extension is to include a new $U(1)$ gauge group, which we call $U(1)^\prime$. This results in two electrically neutral gauge bosons, $Z$ and $Z^\prime$,
that are linear combinations of the SM $Z^0$ boson and the gauge boson of the new $U(1)^\prime$ group. Due to the interconnectedness of the electroweak sector, these additions alter the $W$ boson mass at the tree level which can explain the CDF II measurement.

There are many well-motivated theories beyond the SM that feature at least one extra $U(1)$ group~\cite{Rizzo:2006nw,Langacker:2008yv}, such as grand unified theories~(GUT)~\cite{PhysRevD.26.2396,PhysRevD.34.1530,Preda_2022,Bajc_2007}, superstrings~\cite{Cleaver:1997nj,Cvetic:1999md,PhysRevD.54.3570,Blumenhagen:2005mu}, extra dimensions~\cite{Masip:1999mk}, little Higgs~\cite{Arkani-Hamed:2001nha,Arkani-Hamed:2002iiv,Han:2003wu}, dynamical symmetry breaking~\cite{Hill:2002ap,Hill:2002ap}, and the Stueckelberg mechanism~\cite{Kors:2004iz,Kors:2004ri,Kors:2005uz,Feldman:2006wb,Feldman:2006wd,Cheung:2007ut}. Among the GUT models, the ones based on rank-6 gauge groups, known as $E_6$ 
 have been extensively studied for phenomenological interests~\cite{Hewett:1988xc}. The $E_6$ models can be considered in both supersymmetric and non-supersymmetric scenarios. Extending the Minimal Supersymmetric Standard Model~(MSSM) with an extra $U(1)$ group also has numerous advantages. For example, similar to the Next-to-Minimal Supersymmetric Standard Model~(NMSSM), the tree-level Higgs mass in the $U(1)$-supersymmetric model (UMSSM) is increased, and a 125 GeV Higgs can be obtained without the need of large radiative corrections~\cite{Barger:2006dh}. Furthermore, UMSSM scenarios embed the discrete $Z_3$ symmetry of the NMSSM into a continuous one, and therefore, do not suffer from the cosmological domain walls problems in the NMSSM~\cite{Ellis:1986mq}. 


In this article, we discuss supersymmetric $E_6$ models in light of the CDF II $M_W$ measurement. We note that although our analysis is based on $E_6$, it can easily be generalized to any new physics scenario, supersymmetric or not, with at least one additional $U(1)^\prime$ gauge group. This article is structured as follows. In Sec.~\ref{Sec:mW}, we show the contribution to the $W$ mass from the $U(1)^\prime$ group. In Sec.~\ref{Sec:Constraints} we review the experimental constraints, especially the direct \Zprime searches. These constraints are then applied to $E_6$ models containing the $U(1)^\prime$ group. In Sec.~\ref{Higgs}, we discuss the predictions of the Higgs mass within supersymmetric $E_6$ models. Sec.~\ref{conclusion} is reserved for conclusions. 

\section{Contribution to $M_W$}
\label{Sec:mW}
Models that extend the SM by an extra $U(1)$ gauge group introduce a new gauge boson $Z^\prime$. The Cartan subalgebra of $E_6$ models contains two additional $U(1)$ generators. We consider the following breakdown of $E_6$
\begin{align*}
    E_6 &\rightarrow SO(10)\times U(1)_\psi  \\
    &\rightarrow SU(5) \times U(1)_\chi \times U(1)_\psi \\
    &\rightarrow SU(3)_c \times SU(2)_L \times U(1)_Y \times U(1)_\chi \times U(1)_\psi .
\end{align*}
The two extra $U(1)$ groups yield two additional gauge bosons, $Z_\psi$ and $Z_\chi$. Upon electroweak symmetry breaking, they mix to form two gauge bosons \Zprime and $Z^{\prime\prime}$, with the mixing parameterized by the $E_6$ mixing angle $\theta_{E_6}$,

\begin{align}
    Z^{\prime} &= Z_\chi \cos\theta_{E_6} + Z_\psi \sin\theta_{E_6} \\ \nonumber
    Z^{\prime\prime} &= - Z_\chi \sin\theta_{E_6} + Z_\psi \cos\theta_{E_6}.
\end{align}
Often, only one of the new gauge bosons is assumed to be around the TeV scale, leading to an effective rank-5 group. In this analysis, we will only consider the contributions from the lighter state of the two. We will also allow for a kinetic mixing term, $\frac{\sin{\chi}}{2} B_{\mu}Z^{\prime \mu}$~\cite{Babu96,Chiang:2014yva,Chiang:2015ika,Araz:2017wbp,Araz:2020hyw}, which has been studied in the context of a leptophobic $Z^\prime$. The relevant Lagrangian terms are given in the appendix.

The presence of a new \Zprime boson contributes to the $W$ boson mass at the tree level. The shift in the $W$ boson mass from the SM prediction can be expressed in terms of the oblique parameters $S,T,U$~\cite{Peskin92}: 
\begin{multline} \label{eq:MWO}
    M_W^2 = (M_W^{\text{SM}})^2 \\ + \frac{\alpha c_W^2}{c_W^2 - s_W^2} M_Z^2 \left( -\frac{1}{2} S + c_W^2 T + \frac{c_W^2 - s_W^2}{4s_W^2} U \right)
\end{multline}
where $s_W$ and $c_W$ are the sine and cosine of the weak mixing angle, and $M_Z$ is the physical mass of the SM $Z^0$ boson. The oblique parameters may be derived from the transformation matrix responsible for bringing $\mathcal{L}$ into a basis of fields with canonical kinetic mixing and diagonal mass matrices \cite{Holdom91}. In the appendix we derive this matrix, and from that the oblique parameters.
Here we express the oblique parameters in terms of the mixing angle $\xi$ between the new \Zprime boson and the SM $Z^0$ boson, and the kinetic mixing angle $\chi$ between the $U(1)_Y$ and $U(1)^\prime$ gauge bosons. 
To first order in $\xi$,
\begin{equation} \label{eq:S}
    \begin{split}
        \alpha S
        &= 4 c_W^2 s_W \xi \tan{\chi}, 
    \end{split}
\end{equation}
and $U =0$. $T$ is given by the wavefunction renormalization $\Delta_{Z}$ of the $Z$ boson (found in the transformation matrix) as well as the shift in the $Z$ boson mass from its SM prediction. To first order in $\xi$, the wavefunction renormalization is
\begin{equation}
    \Delta_Z = s_W \xi \tan\chi.
\end{equation}

With $Z-$\Zprime mass mixing, the tree-level $Z$ boson mass $M_Z$ is shifted from its SM value $m_Z$ as
\begin{equation} \label{eq:identity}
    m_Z^2 =M_Z^2 + [M_{Z'}^2 - M_Z^2]\sin^2{\xi}
\end{equation}
which is an identical relation between the $Z-$\Zprime mass matrix and its diagonalized form. For small $Z-$\Zprime mixing angles of $\xi^2 \ll M_Z^2 / M_{Z'}^2$, Eq. \eqref{eq:identity} is approximately
\begin{equation}
    M_Z^2 \approx m_Z^2 \left[ 1 - \xi^2 \left( \frac{M_{Z^\prime}^2}{M_Z^2} - 1 \right) \right].
\end{equation}

These changes to the properties of the $Z$ boson are combined to form the $T$ parameter:
\begin{equation} \label{eq:T}
\begin{split}
    \alpha T &= 2( \Delta_Z - \Tilde{\Delta}_Z ) \\
    &= 2\xi s_W \tan{\chi} + \xi^2 \left( \frac{M_{Z^\prime}^2}{M_Z^2} - 1 \right)
\end{split}
\end{equation}
where $\Tilde{\Delta}_Z$ is the fractional mass shift of $Z$ from the SM, given to first order in $\xi$.

\begin{figure}[h]
    \includegraphics[scale=0.6]{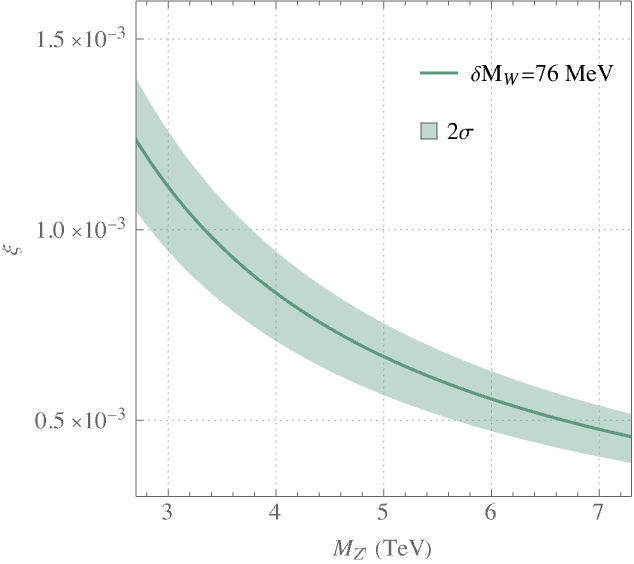}
    \caption{The solid line shows the solution to Eq.~(\ref{eq:MW}) for $\delta M_W = 76$~MeV, which is the central value of the deviation of the SM from the CDF II experiment. The shaded region contains solutions for $54\,\, \text{MeV} \leq \delta M_W \leq 98 \,\,\text{MeV}$, corresponding to a 2$\sigma$ confidence level of the CDF II measurement.  }
    \label{fig:Wmass}
\end{figure}

Neglecting terms of order $\delta M_W^2$, Eq.~\eqref{eq:MWO} now yields the following $W$ boson mass shift:
\begin{equation} \label{eq:MW}
    \delta M_W \approx \frac{1}{2 M_W} \frac{c_W^4}{c_W^2 - s_W^2} \xi^2 \left( M_{Z^\prime}^2 - M_Z^2 \right),
\end{equation}
which only depends on the \Zprime mass and $Z-$\Zprime mixing. In Fig.~\ref{fig:Wmass}, we plot the solution to Eq.~(\ref{eq:MW}) in the $M_{Z^{\prime}} - \xi$ plane. 

\section{Experimental Constraints}
\label{Sec:Constraints}
In this section, we will discuss various experimental constraints on a new $Z^\prime$ boson. The \Zprime may be directly produced at the Large Hadron Collider (LHC) through $p \,p \rightarrow Z^{\prime} \rightarrow l\,l$ processes, and is therefore subject to current resonant dilepton searches~\cite{CMS13,ATLAS:2017fih}.
The results are presented as the upper limit on the product of the \Zprime production cross section $\sigma$ with the branching ratio of the \Zprime to 
dilepton pairs for various \Zprime masses, $\sigma (p\, p \rightarrow Z^{\prime}) \times BR(Z^{\prime} \rightarrow l^+l^-)$. Further, the results are used to constrain the Sequential Standard Model~(SSM), and some rank-5 scenarios in the $E_6$ model. In general $E_6$ models, with a possible kinetic mixing term and $Z-Z^{\prime}$ mixing, the neutral current $J^\mu_2$ of the heavier mass eigenstate can be written as
\begin{equation}
    J^\mu_2 = \sum_i g_{Z^{\prime}} \Bar{f}_i \gamma^\mu [Q_{2iL} P_L + Q_{2iR} P_R] f_i,
\end{equation}
where $g_{Z^{\prime}}$ is the coupling constant of the new gauge group. If we assume grand unification, $g_{Z^{\prime}} = 0.46$ at the electroweak scale~\cite{Langacker:2008yv}. $Q_{2i(L,R)}$ are found in the appendix to be 
\begin{equation} \label{eq:Z2charge}
\begin{split}
Q_{2i(L,R)} =& \frac{g_1}{g_{Z^{\prime}}}c_W [-c_W \cos{\xi} \tan{\chi}] q_i \\
    & + \frac{g_2}{c_W g_{Z^{\prime}}} [-\sin{\xi} + s_W \cos{\xi} \tan{\chi}] Q_{Zi(L,R)}  \\
        & + [\cos{\xi} / \cos{\chi}] Q^\prime_{i(L,R)}.
\end{split}
\end{equation}
The first two terms are contributions from the photon and SM $Z^0$ boson components in the new \Zprime boson due to the mixing. $g_1$ and $g_2$ are the coupling constants of the SM $U(1)_Y$ and $SU(2)_L$ gauge groups. $q_i$ is the electromagnetic charge of fermion $f_i$, and \begin{equation} \label{eq:Zcharges}
    Q_{Zi(L,R)} = T^3_{i(L,R)} - q_i s_W^2,
\end{equation}
where $T^3_{i(L,R)}$ is the third isospin component of fermion $f_i$. The last term in Eq.~\eqref{eq:Z2charge} is the contribution due to the new $U(1)^{\prime}$ gauge group and $Q^\prime_{i(L,R)}$ are the charges of fermions under this group. 

As noted above, within $E_6$ models the $U(1)^\prime$ group is taken as an orthogonal mixture of groups $U(1)_\chi$ and $U(1)_\psi$ such that the group generators $Q'$, $Q_\chi$, and $Q_\psi$ are related through
\begin{equation} \label{eq:E6}
    Q^\prime = Q_\chi \cos{\theta_{E_6}} + Q_\psi \sin{\theta_{E_6}}.
\end{equation}
The charges for the fermions are listed in Table~\ref{tab:E6_charges}.

Table~\ref{tab:E6} lists canonical $E_6$ models which are anomaly-free without the requirement of additional massless fields. However, anomalies are present in models for all other $\theta_{E_6}$ values. 
In those cases, to cancel the anomalies, one needs to introduce the complete multiples~\cite{Roepstorff:2000up}. Those lighter states can also contribute to $M_W$ through loop effects. In this article, we only focus on the tree-level effects. Those models also contain right-handed neutrinos $\nu_R$ for anomaly cancellation. With the right-handed neutrinos, one can generate small neutrino masses through the $Y \overline{\nu}_R L H_u+$ h.c. interaction, where $L$ is the SM leptonic doublet. However, with the \Zprime around the TeV scale, unless the right-handed neutrinos carry a zero $U(1)^{\prime}$ charge, as in the $N$ model, they cannot obtain the large Majorana mass needed for the conventional seesaw mechanism. In this case, small Dirac or Majorana neutrino masses are possible by invoking alternatives to the conventional seesaw mechanisim~\cite{Langacker:2008yv}.
\begin{table}[h]
    \centering
    \begin{tabular}{|c||c|c|}
    \hline
         Field & $2 \sqrt{10} Q_\chi$ & $2 \sqrt{6} Q_\psi$ \\
         \hline
         $u_L$ & $-1$ & 1 \\
         $u_R$ & 1 & $-1$ \\
         $d_L$ & $-1$ & 1 \\
         $d_R$ & $-3$ & $-1$ \\
         $e_L$ & 3 & 1 \\
         $e_R$ & 1 & $-1$ \\
         $\nu_L$ & 3 & 1 \\
         $\nu_R$ & 5 & $-1$ \\
         \hline
         $H_u$ & 2 & $-2$ \\
         $H_d$ & $-2$ & $-2$ \\
         $S$ & 0 & 4 \\
         \hline
    \end{tabular}
    \caption{$E_6$ charges for fermion and scalar fields following Ref.~\cite{Langacker:2008yv}.}
    \label{tab:E6_charges}
\end{table}

\begin{table}[h]
    \centering
    
    \begin{tabular}{|c ||c|}
    \hline
    Model & $\theta_{E_6}$ \\
    \hline
    $\chi$ & 0 \\
    $\psi$ & $\pi/2$ \\
    $\eta$ & $- \tan^{-1}{\sqrt{5/3}}$ \\
    $I$ & $\tan^{-1}{\sqrt{3/5}}$ \\
    $N$ & $\tan^{-1}{\sqrt{15}}$\\
    $S$ & $\tan^{-1}{\sqrt{15}/9}$\\
    \hline
    \end{tabular}
     \caption{Canonical examples of $E_6$ models.}
    \label{tab:E6}
\end{table}


In this analysis, we recast cross-sectional bounds found by the CMS experiment by considering a ratio of cross sections in $E_6$ models and in the SSM. We parameterize the ratio following Ref.~\cite{Barger13},
\begin{equation} \label{eq:ratio}
    \frac{\sigma(pp \rightarrow Z^\prime \rightarrow l^- l^+)}{\sigma(pp \rightarrow Z^\prime_{\text{SSM}} \rightarrow l^- l^+)} =
    \frac{p c_u + (1-p)c_d}{p c_{u, \text{SSM}} + (1-p)c_{d, \text{SSM}}}.
\end{equation}
The parameter $p$ is a numerical fit depending on $M_{Z^\prime} / \sqrt{s}$ to account for the parton distribution function:
\begin{equation} \label{eq:p}
    p = 0.77 - 0.17 \tan^{-1}\left(2.6 - 9.5 \frac{M_{Z^\prime}}{\sqrt{s}} \right)
\end{equation}
with $\sqrt{s}$ being the center-of-momentum energy. The production cross section and branching ratios are written with quantities $c_u$, $c_d$ where
\begin{align}
    c_q &= \frac{M_{Z^\prime} g_{Z^{\prime}}^4}{24 \pi \Gamma_{Z^\prime}} (Q_{2qL}^{2} + Q_{2qR}^{2}) (Q_{2eL}^{2} + Q_{2eR}^{2}) \label{eq:c} \\
    c_{q,SSM} &= \frac{M_{Z^\prime} g_{Z}^4}{24 \pi \Gamma_{Z^\prime}} (Q_{ZqL}^2 + Q_{ZqR}^2) (Q_{ZeL}^2 + Q_{ZeR}^2). \label{eq:cSSM}
\end{align}
Here, $\Gamma_{Z^\prime}$ is the full width of the $Z^\prime$ boson found by summing partial widths for decays into massless fermion-antifermion pairs. In $E_6$ models,
\begin{equation} \label{eq:width}
    \Gamma_{Z^\prime} = \sum_f g_{Z^{\prime}}^2 \frac{M_{Z^\prime}}{24 \pi} (Q_{2fL}^2 + Q_{2fR}^2),
\end{equation}
assuming negligible fermion masses.
 In the SSM, the couplings $g_{Z^\prime}$ and charges $Q_{2i(L,R)}$ are replaced by $g_Z \equiv g_2 /c_W$ and $Q_{Zi(L,R)}$ respectively. When calculating the \Zprime width, we assume the new $E_6$ fermions are heavy enough to be ignored.

To establish a parameter space, we calculate the ratio of Eq.~\eqref{eq:ratio} and compare with its upper bound set by CMS searches at the LHC at $\sqrt{s}=13$ TeV, as well as CMS projections of high luminosity LHC (HL-LHC) at $\sqrt{s}=$ 14 TeV \cite{CMS13,CMS14}.
The production and decay of \Zprime depends on its mass $M_{Z^\prime}$ and the charges $Q_{2i (L,R)}$ contributing to the neutral current $J_2^{\mu}$. As shown in Eq.~\eqref{eq:Z2charge}, $Q_{2i (L,R)}$ are determined by the $Z-$\Zprime mixing $\xi$, kinetic mixing $\chi$, and the $E_6$ mixing angle $\theta_{E_6}$.
\begin{figure}[tbh!]
     \centering
     \begin{subfigure}[b]{0.33\textwidth}
         \centering
         \includegraphics[width=\linewidth]{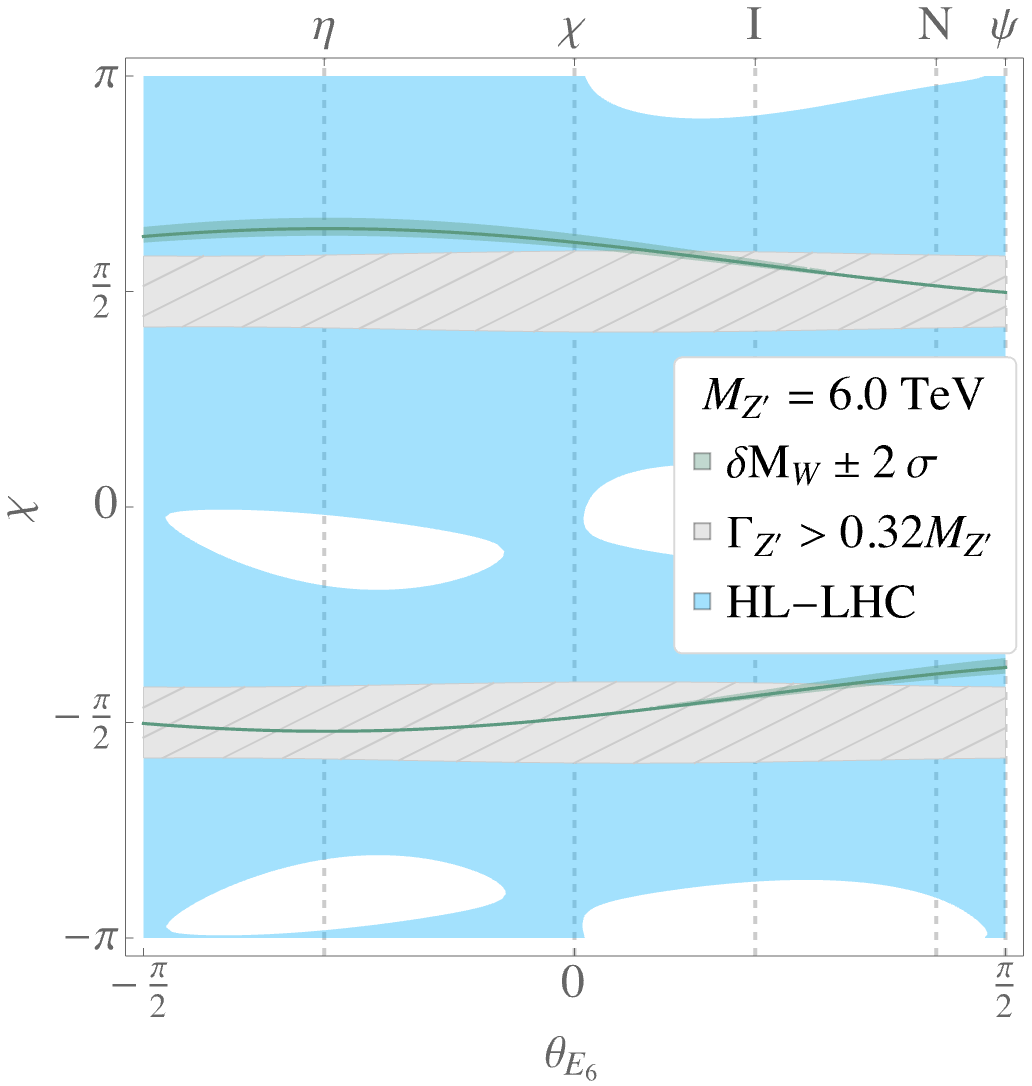}
         \label{fig:5500}
     \end{subfigure}
     \hfill
     \begin{subfigure}[b]{0.33\textwidth}
         \centering
         \includegraphics[width=\linewidth]{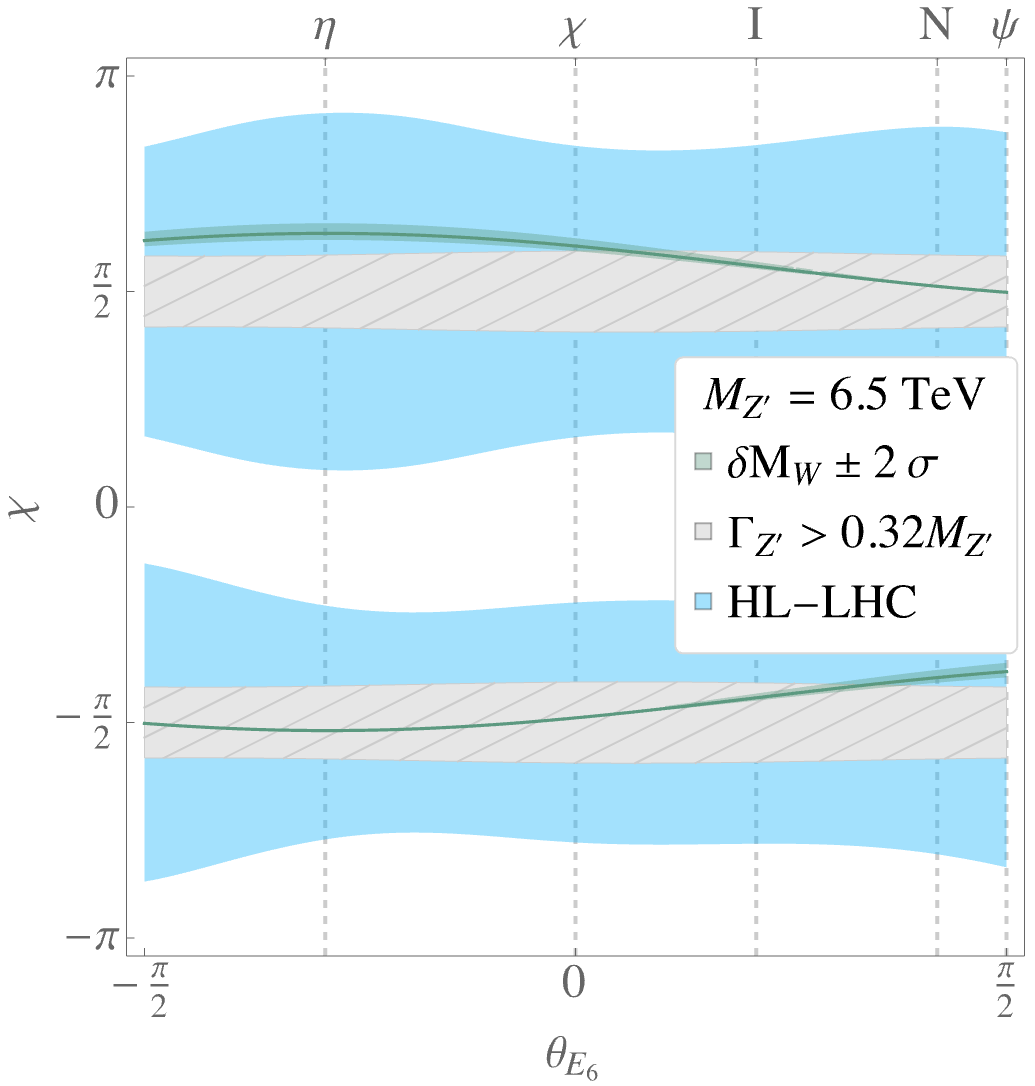}
         \label{fig:6000}
     \end{subfigure}
     \hfill
     \begin{subfigure}[b]{0.33\textwidth}
         \centering
         \includegraphics[width=\linewidth]{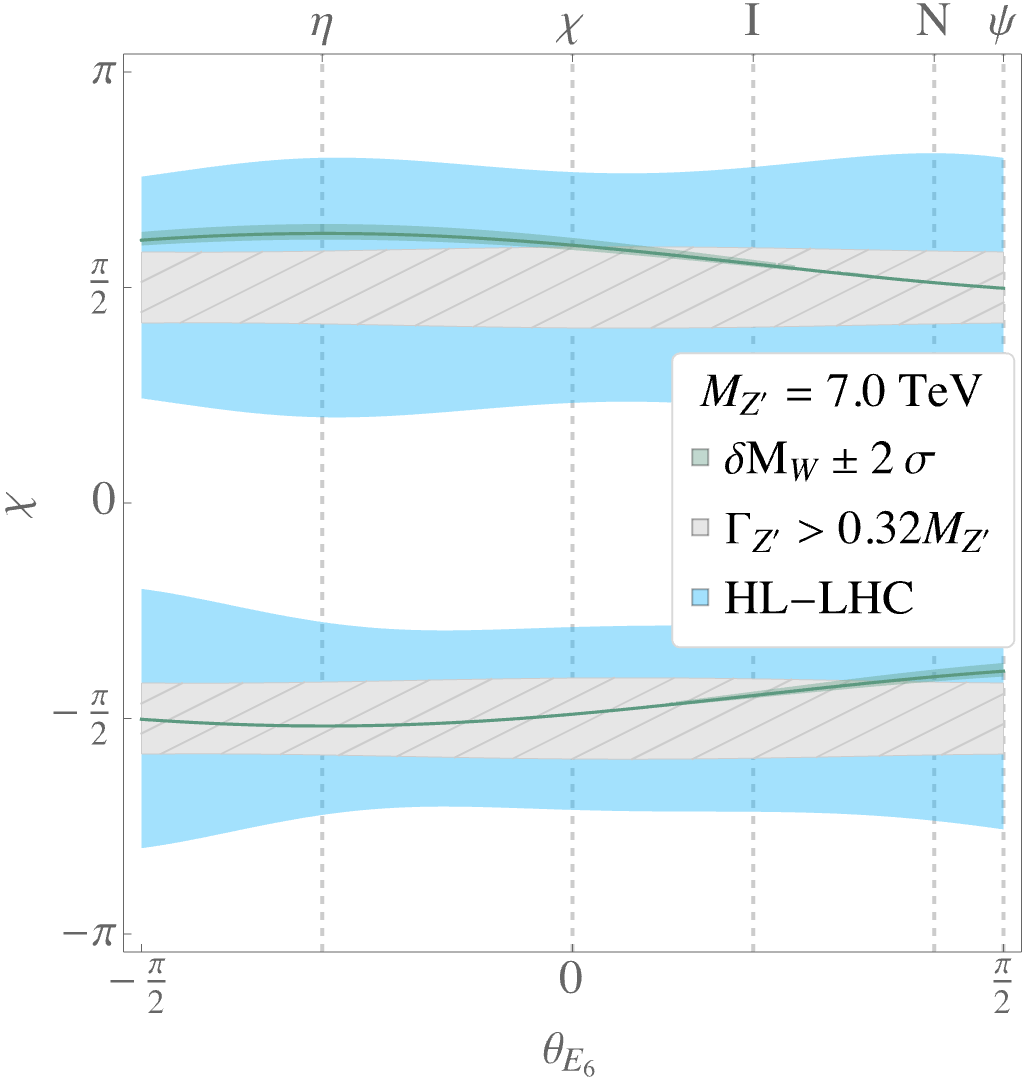}
         \label{fig:6500}
     \end{subfigure}
        \caption{Parameter space probeable by HL-LHC resonant dilepton searches at 14 TeV (blue) and a wide resonance region of $\Gamma_{Z^\prime}/M_{Z^\prime}>32\%$ (gray) for $M_{Z^{\prime}} =
    $ 6~TeV (top), 6.5~TeV (middle), and 7~TeV (bottom). Regions consistent with the CDF II $W$ mass measurement are shown in green.
        }
        \label{fig:CMS}
\end{figure}
Additionally, Eq.~\eqref{eq:MW} fixes $\xi$ for a given $M_{Z^\prime}$ to satisfy the CDF II result Eq.~\eqref{eq:CDF}, and $\xi$ is determined by $\chi$ and $M_{Z^{\prime}}$, reducing the parameter space by two (using Eq.~\eqref{eq:mixing} in the appendix and $\tan\beta=10$). The $\sqrt{s}=13$~TeV search excludes $Z^\prime$ models that explain the CDF II result with $M_{Z^\prime} \leq 5.5$~TeV. At higher masses, the resulting space which may be probed by the $\sqrt{s}=14$~TeV CMS search is shown in Fig. \ref{fig:CMS} by the blue region.


There exists open parameter space around $|\chi| = \pi/2$ due to the diverging behavior of both $M_{Z^\prime}$ and the chiral couplings $Q_{2 i (L,R)}$ in this region. As the mass and couplings increase in magnitude, so does the full width $\Gamma_{Z'}$ of the \Zprime boson. This widening of $\Gamma_{Z'}$ makes it easier for the \Zprime to evade direct LHC searches. 
We close off these regions with gray hashing in Fig.~\ref{fig:CMS} where the total \Zprime width is more than 32\% of the \Zprime mass following the ATLAS study in~\cite{ATLAS:2017fih}. This study finds that wide resonances with $\Gamma_{Z^\prime} / M_{Z^\prime} \leq 32\%$ do not significantly affect search bounds utilizing the narrow width approximation (NWA) for $Z^\prime$s heavier than 4.5~TeV, and those effects become less significant as $M_{Z^\prime}$ increases \cite{CMS13,ATLAS:2017fih}. Even so, it should be noted that our use of the NWA introduces estimated errors of $\mathcal{O}(\Gamma_{Z^\prime} / M_{Z^\prime})$ \cite{Berdine:2007uv}. Interference effects from SM gauge boson production also influence the sensitivity of collider searches. We do not consider these effects, however relative interference at the LHC can be as low as a few percent for searches that assume narrow widths \cite{Accomando:2010fz}.

For a sufficiently light $Z^\prime$, holes may be found in the probeable parameter space; as shown in the top panel of Fig.~\ref{fig:CMS}. There are two different charge suppressions responsible for the holes near the $I$ and $\eta$ models. For holes along $\theta_{E_6}$ near the $I$ model, these regions maintain small \Zprime charges for up quarks which suppress \Zprime production from proton collisions. On the other hand, the regions near holes along the $\eta$ model maintain small \Zprime charges for leptons which in turn yield small production cross sections in the lepton channels. These findings are consistent with leptophobic studies within the $\eta$ model \cite{Babu96,Chiang:2014yva}. In either case, the $pp \rightarrow Z^\prime \rightarrow l^- l^+$ production inside these holes is small enough to evade the cross-sectional upper bounds found by CMS.

In addition to direct searches, a \Zprime gives rise to various corrections to the properties of the $Z$ boson through $Z-Z^{\prime}$ mixing parameter $\xi$, which is tightly constrained by precision $Z$ boson measurements. As shown in Fig.~\ref{fig:Wmass}, the $Z-Z^{\prime}$ mixing required to explain the $W$ mass measurement is well below $10^{-3}$ for $Z^\prime$ bosons heavier than 4~TeV. The combined fit for $Z-$pole observables put an upper bound on the $Z-Z^{\prime}$ mixing parameter around $3\times10^{-3}$~\cite{Babu97,Erler:1999nx,Ferroglia:2006mj,Alguero_2022}. We have checked that the kinetic mixing introduced here did not lead to modifications beyond the current precision. For instance, a 6 TeV \Zprime that explains the current $W$ mass measurement in the $\eta$ model has a deviation in the leptonic decay width of the $Z$ boson of 0.0029\%, which is within experimental uncertainties~\cite{PDG}. 

\section{Higgs Mass} \label{Higgs}

In SUSY models, the mass $m_h$ of the Higgs boson is precisely predicted by a few relevant parameters, and can be calculated through fixed-order calculations, effective field theory~(EFT) calculations, and a hybrid calculation. Dominant three-loop contributions to $m_h$ are known in the MSSM.~(For $m_h$ calculation in SUSY models, see~\cite{Slavich:2020zjv} and references therein.) At the tree-level, the Higgs mass has an upper bound of $M_Z \cos 2\beta$ in the MSSM. It receives substantial radiative corrections with the dominant contributions coming from loops involving the top, and its scalar partner, the stop, along with gluon and gluino exchanges. In particular, when the SUSY scale $M_S$ is around 2~TeV and the stop mixing parameter is $X_t \simeq -\sqrt{6} M_S$, a 125~GeV Higgs can be achieved with $|H^0_u|/|H^0_d| \equiv \tan\beta \gtrsim 10 $ where $|H^0_u|$ is the vacuum expectation value (VEV) of the electrically neutral component of the doublet Higgs field $H_u$. $M_S$ is predicted to be at least 10~TeV from $m_h = 125$~GeV with a vanishing stop mixing parameter~\cite{Slavich:2020zjv}. The current theoretical uncertainties in calculating $m_h$ are around 2-3~GeV for the MSSM~\cite{Slavich:2020zjv}.  

We propose extending the SM gauge group to explain the latest $M_W$ measurement. Expanding into SUSY scenarios, when the MSSM is extended by an extra $U(1)$ group there are additional contributions to the $D$-term~\cite{Barger:2006dh},
\begin{equation}
    \frac{g_{Z^{\prime}}^2} {2}(Q^{\prime}_{H_d} |H_d^{0}|^2  +Q^{\prime}_{H_u} |H_u^{0}|^2 + Q^{\prime}_{s} |S|^2 )^2, 
\end{equation}
in which $H_u$ and $H_d$ are the Higgs doublets from the MSSM, and $S$ is the singlet scalar field that breaks the new $U(1)^\prime$ symmetry. $Q^{\prime}_{H_u}$, $Q^{\prime}_{H_d}$, and $Q^{\prime}_s$ are their corresponding charges under the $U(1)^\prime$ group. We impose the charge-conserving relation $Q^{\prime}_{H_u} + Q^{\prime}_{H_d} + Q^{\prime}_{s} = 0$, coming from the $\lambda S H_u H_d$ term in the superpotential. $E_6$ charges for the scalar fields are given in Table~\ref{tab:E6_charges}. In addition to the new $D$-term contribution, the $\lambda S H_u H_d$ term in the superpotential increases the upper bound of the Higgs mass as in the NMSSM~\cite{Maniatis:2009re,Ellwanger:2009dp}. Combining both contributions, the upper bound of the Higgs mass becomes~\cite{Barger:2006dh}
 
\begin{equation}
\begin{split}
    m_h^2 =& M_Z^2 \cos^2 2\beta + \lambda^2 v^2 \sin^2 2\beta \\
    & + g_{Z^{\prime}}^2 v^2 (Q^{\prime}_{H_d}\cos^2\beta + Q^{\prime}_{H_u}\sin^2\beta)^2.
    \label{eq:mh}
\end{split}
\end{equation}
The increased tree-level Higgs mass implies that the stop sectors are much less constrained in the MSSM case.

To account for loop effects and the effects of the running couplings, we use FlexibleSUSY for our numerical analysis. FlexibleSUSY~\cite{FlexibleSUSY2015,FlexibleSUSY2018} is a Mathematica and C++ package for generating mass spectra of SUSY models. It includes 2-loop Renormalization Group Equations (RGEs), it calculates $m_h$ at the full 1-loop level, and it includes dominant corrections up to 3-loop, next to leading logarithms. The theoretical uncertainty in the UMSSM scenario calculated in FlexibleSUSY was estimated to be as large as $\pm$10 GeV~\cite{Athron:2016fuq}. The large uncertainty compared to the MSSM case is due to the altered RGEs in the $E_6$ scenarios.   
For the MSSM-like parameters, we adopt a benchmark point motivated by the Natural SUSY scenario~\cite{Baer_2022}, in which $m_0 = 7.9$~TeV, $m_{1/2} = 1.2$~TeV, $\tan\beta = 10$, $A_0 = -8$~TeV, $\mu_{\text{eff}} = 200$~GeV, and $m_A=2$~TeV. With the new $U(1)^\prime$ gauge group, there are two more free parameters: the gauge coupling $g_{Z^{\prime}}$ which we fix to be 0.46 from grand unification, and the VEV of the singlet field $S$, $|S| = \sqrt{2} \mu_{\text{eff}}/ \lambda$ .
For some $\theta_{E_6}$, $M_{Z^\prime}$, and kinetic mixing $\chi$, $|S|$ is specified, and therefore $\lambda$ is specified. As seen in Fig.~\ref{fig:CMS}, for a given $Z^{\prime}$ mass and $\theta_{E_6}$, there are two solutions in the range $-\pi \leq \chi \leq \pi$ that explain the new $W$ mass measurement. We choose the solution for which $|\chi| - \pi/2$ is maximized to avoid the diverging full width of the $Z'$ boson at $|\chi| = \pi/2$.

\begin{figure}[tbh!]
\begin{subfigure}[b]{0.4\textwidth}
    \includegraphics[width=\linewidth]{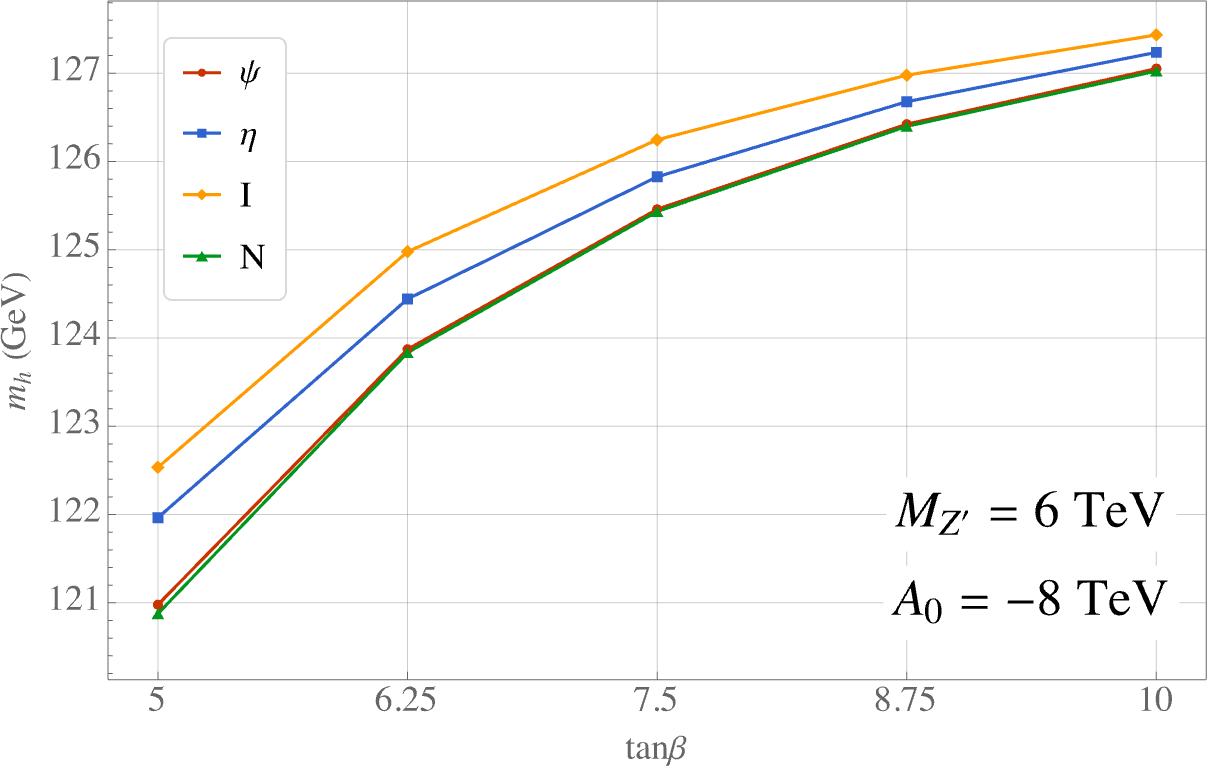}
\end{subfigure}
\hfill
\caption{Masses $m_h$ of the lightest scalar mass eigenstate in different $E_6$ models for the benchmark point in Section \ref{Higgs} ($m_0 = 7.9$ TeV, $m_{1/2} = 1.2$ TeV, $A_0 = -8$ TeV, $\mu_{\text{eff}} = 200$~GeV, $m_A=2$ TeV) which satisfy the CDF II measurement.}
\label{fig:higgs}
\end{figure}

Shown in Fig.~\ref{fig:higgs} are predictions calculated by FlexibleSUSY for $m_h$, the mass of the lightest mass eigenstate within the model's scalar sector. 
We fix $M_{Z^{\prime}} = 6$~TeV and vary $\tan\beta$. As expected, $m_h$ is increased compared to the MSSM case (125 GeV for the benchmark we chose), and it is reduced as $\tan\beta$ decreases. In particular, for all $E_6$ scenarios we consider, $m_h$ is about 125 GeV for $\tan\beta \simeq$ 7 in this benchmark. 
Conversely, $m_h$ depends very weakly on $M_{Z^{\prime}}$ due to the following. The dependence of $m_h$ on $M_{Z^{\prime}}$ is through $\lambda = \sqrt{2} \mu_{\text{eff}}/|S|$, which is suppressed by $\sin^2 2\beta$. Furthermore, $M_{Z^{\prime}}$ only depends on $|S|$ weakly. The dominant contribution to the \Zprime mass is
\begin{equation}
  M_{Z{^{\prime}}}^2 \simeq g_{Z^{\prime}}^2 (Q_{H_u}^{\prime 2} v^2 \sin^2 \beta +Q_{H_d}^{\prime 2} v^2 \cos^2 \beta + Q_{s}^{\prime2}|S|^2)/\cos^2\chi
  \label{eq:MPZapprox}
\end{equation}
   (the full result is presented in the appendix). We found that heavier a \Zprime requires a larger kinetic mixing $\chi$ to explain the CDF II $W$ mass measurement. This increase in $\chi$ yields a heavier \Zprime without requiring a large $|S|$. 
   
   Results for the $\chi$ model are not shown in Fig.~\ref{fig:higgs} because $Q_s^\prime = 0$ in this model. Eq.~(\ref{eq:MPZapprox}) shows that a heavy \Zprime can not be achieved with a vanishing $Q_s^{\prime}$ unless the gauge coupling $g_Z^{\prime}$ is very large. In the above numerical analysis, we do not include the kinetic mixing contribution to the $D$-term in the $m_h$ calculation. We have checked that it can lead to an up to $\pm$2~GeV shift in the Higgs mass at the tree-level. As discussed, when we adopt the benchmarks from the Natural SUSY scenarios, in general, the predicted $m_h$ is larger than 125~GeV. With the same set of parameters, we found the Higgs mass to be closer to 125 GeV with $\tan\beta \gtrsim 10$ and $A_0 \simeq 0$ across all $E_6$ models discussed in this work. The possibility to accommodate a 125~GeV Higgs with small mixing in the stop sector is an encouraging feature, as the stop mixing is naturally small in minimal Anomaly mediated SUSY models~\cite{Randall:1998uk,Giudice:1998xp,Pomarol:1999ie,Bagger:1999rd,Jack:1997eh,Gaillard:2000fk,Binetruy:2000md,Dine:2007me,deAlwis:2008aq,Dine:2013nka,PhysRevD.98.015039} and Gauge mediated SUSY models~\cite{Dine:1981gu,Dine:1981rt,Dimopoulos:1981au,Alvarez-Gaume:1981abe,Nappi:1982hm,Eu:2021sig,Everett:2018wrn}. 

\begin{table}[hb]
    \centering
    \begin{tabular}{|c||c|c|c|c|}
    \hline
         & BM1 ($\eta$) & BM2 ($\eta$) & BM3 ($N$) & BM4 ($\psi$) \\
        \hline
        $\theta_{E_6}$ (rad) & $-0.91$ & $-0.91$ & $1.32$ & $1.57$ \\
        $M_{Z^{\prime}} (\text{TeV})$ & 6 & 6.5 & 6.5 & 7 \\
        
        $A_0$ (TeV)& 0 & $-2$ & $0$ & $-8$ \\
        $\tan\beta$ & 50 & 8 & 10 & 7 \\
        \hline
    $\chi$ (rad) & $2.03$ & $1.99$ & $-1.24$ & $-1.23$ \\
        $m_{\tilde{t}_1}$ (TeV) & 4.86 & 4.76 & 4.65 & 3.39 \\
        $m_{\tilde{\chi}_1^\pm} $ (GeV) & 193 & 244 & 253 & 254 \\ 
        $m_{\tilde{\chi}_1^0}$ (GeV) & 187 & 236 & 245 & 247 \\
        $m_{\tilde{\chi}_2^0}$ (GeV) & 198 & 250 & 259 & 261 \\
        
        $m_h$ (GeV) & 126.51 & 125.20 & 126.03 & 124.95 \\
        \hline
        $BR(Z^\prime \rightarrow l^+ l^-) $ & 16.0\% & 16.1\% & 10.5\% & 12.5\% \\
        $\Gamma_{Z^\prime} / M_{Z^\prime} $ & 10.0\% & 12.1\% & 20.7\% & 18.1\% \\
        \hline
    \end{tabular}
    \caption{
    Four benchmark points with predicted $M_W$ consistent with the CDF II measurement. For all four points, $m_0 = 7.9$~TeV, $m_{1/2} = 1.2$~TeV, $\mu_{\text{eff}} = 200$~GeV, and $m_A=2$~TeV.
   }
    \label{tab:masses}
\end{table}
   
   In Table~\ref{tab:masses}, we present four benchmark points motivated by the Natural SUSY scenario with various \Zprime masses. For each point, the $W$ mass is at the central value of the CDF II measurement, and the \Zprime mass is allowed by current CMS results. We provide in Table~\ref{tab:masses} other SUSY parameters, the Higgs mass, and the particle spectrum of the benchmark points. For all four benchmarks, the gluino mass is 3.03~TeV, which is near the projected sensitivity of the HL-LHC~\cite{ATL-PHYS-PUB-2014-010,CMS-PAS-FTR-13-014}. In the first three benchmarks, a Higgs mass near 125~GeV is achieved by having a small $A_0$. 
   In this case, BM1 shows that a reasonable Higgs mass can be achieved with large $\tan\beta$. BM4 represents the possibility of accommodating a 125~GeV Higgs by having a small $\tan\beta$ and large mixing in the stop sector. 
   For the electroweakino sector, as in the Natural SUSY scenarios, the lightest chargino and the two lightest neutralinos are Higgsino-like. Those electroweakinos are produced with sizable rates at the LHC and can be searched for with a soft dimuon trigger, or a hard initial state radiation jet, or through the mono-jet channel~\cite{Giudice:2010wb,Gori:2013ala,PhysRevD.90.115007,Han:2014kaa,Han:2015lma,Baer:2016usl,Baer:2020sgm,Arganda:2021lpg}. The Higgsinos may also be accessible at lepton colliders~\cite{Baer:2011ec}.
   Additionally, widths and leptonic branching ratios for the \Zprime are listed at the bottom of Table~\ref{tab:masses} with $l$ being an electron or a muon.


\section{Conclusion} \label{conclusion}
In summary, we point out a tree-level contribution to the $W$ mass in supersymmetric $E_6$ models with kinetic mixing. 
The precision of the latest CDF II mass measurement of the $W$ boson tightly restricts $Z-$\Zprime mixing to be $\xi \sim O(10^{-3})$ for \Zprime masses in the TeV range. When combined with the direct dilepton resonance searches at the LHC, further constraints are placed on the $E_6$ models. For example, we have checked that for $M_{Z^\prime}\leq 5.5$~TeV, $\theta_{E_6}$ models are excluded at the $2\sigma$ level by the CDF II measurement and CMS searches at $\sqrt{s}=13$ TeV \cite{CMS13}. Moreover, we show how the HL-LHC run at 14~TeV is projected to further probe $E_6$ models for more massive \Zprime bosons. Additional calculations are made for the Higgs boson mass within a UMSSM. We find that a 125~GeV Higgs is possible within the reasonable parameter space allowed by experiments. 

As for future directions, it will be interesting to study the associated phenomenology of the SUSY particles. With precision $W$ boson and Higgs mass measurements, the scale and mixing of the stop sector can be predicted. 
Motivated by the natural SUSY scenario (in which the higgsinos are light) we also expect rich phenomenology in the electroweakino sector. Thus, there is a complementarity between direct \Zprime searches, direct stop and electroweakino searches, and precision $M_W$ and $m_h$ measurements as they work together towards revealing new physics beyond the SM.
\acknowledgments 
The work of VB is supported by the U.S. Department of Energy, Office of Science, Office of High Energy Physics under Grant DE-SC-0017647. The work of CH and PH is supported by the National Science Foundation under grant numbers PHY-1820891 and PHY-2112680, and the University of Nebraska Foundation.  
\appendix

\section{}
\subsection{Lagrangian} \label{appendix:Lagrangian}
There are three scalar fields in our model: two Higgs doublets from the MSSM, and a singlet scalar field that breaks the new $U(1)^\prime$ symmetry. Their $SU(2)_L \times U(1)_Y \times U(1)^\prime$ group representations are
\begin{align*}
H_u &\sim (\textbf{2}, 1/2, Q^\prime_{H_u}) \\
H_d &\sim (\textbf{2}, -1/2, Q^\prime_{H_d}) \\
S &\sim (\textbf{1}, 0, Q^\prime_S).
\end{align*}
The covariant derivative is
\[ D_\mu = \partial_\mu -i[g_2 W^a_\mu T^a + g_1 B_\mu Y + g_{ Z^\prime} Q^\prime ] \]
where: $W^a_\mu$, $B_\mu$, $Z^\prime_\mu$ are the $SU(2)_L$, $U(1)_Y$, $U(1)^\prime$ gauge bosons; $T^a$, $Y$, $Q^\prime$ are the group generators; $a$ runs from 1 to 3.

After the electrically neutral components of the scalar fields acquire VEVs, their kinetic terms $(D_\mu H)^\dagger D^\mu H$ in the Lagrangian will yield mass terms for vector fields $Z_\mu=c_W W^3_\mu - s_W B_\mu$ and $Z_\mu^\prime$ shown below.  The weak mixing angle $\theta_W$ is defined by $s_W / c_W = \sin\theta_W / \cos\theta_W \equiv g_1 / g_2$. The SM values for these parameters \cite{PDG} are used throughout this study.

The relevant Lagrangian terms are given by

\[
\mathcal{L} \supset \mathcal{L}_{\text{mass}} + \mathcal{L}_{\text{kinetic}} + \mathcal{L}_{\text{current}},
\]
\begin{equation*}
    \mathcal{L}_{\text{mass}} = -\frac{1}{2} m_Z^2 Z_\mu Z^\mu - \frac{1}{2} m_{Z^\prime}^2 Z^\prime_\mu Z^{\prime \mu} - \Delta^2 Z_\mu Z^{\prime \mu}
\end{equation*}
\begin{equation*}
\begin{split}
    \mathcal{L}_{\text{kinetic}} =& -\frac{1}{4} W^3_{\mu \nu} W^{3 \mu \nu} -\frac{1}{4} B_{\mu \nu} B^{\mu \nu} -\frac{1}{4} Z^\prime_{\mu \nu} Z^{\prime \mu \nu} \\ &-\frac{\sin{\chi}}{2} B_{\mu \nu} Z^{\prime \mu \nu}
\end{split}
\end{equation*}
\begin{equation*}
    \mathcal{L}_{\text{current}} = - J_3^\mu W^3_\mu - J_Y^\mu B_\mu - J^{\prime \mu} Z^\prime_\mu
\end{equation*}
where $F_{\mu \nu}$ is the field strength tensor for a gauge field $F_\mu$. The fermion currents are
\begin{align}
    J^\mu_3 &= g_2 \sum_i \Bar{f}_i \gamma^\mu [T^3_{iL} P_L + T^3_{iR} P_R] f_i \\
    J^\mu_Y &= g_1 \sum_i \Bar{f}_i \gamma^\mu [Y_{iL} P_L + Y_{iR} P_R] f_i \\
    J^{\prime \mu} &= g_{Z^\prime} \sum_i \Bar{f}_i \gamma^\mu [Q^\prime_{iL} P_L + Q^\prime_{iR} P_R] f_i.
\end{align}
A group generator indexed by $iL$ ($iR$) gives the corresponding charge of the left-handed (right-handed) component of fermion field $f_i$. $P_L$ and $P_R$ are the left- and right-projection operators.

The terms in $\mathcal{L}$ which are absent from the SM all involve the new $Z'_\mu$ gauge field. In particular, $\mathcal{L}_{\text{kinetic}}$ and $\mathcal{L}_{\text{mass}}$ contain mixing terms which must be negotiated to find the proper mass eigenstates within $\mathcal{L}_{\text{current}}$.

\subsection{Kinetic Mixing} \label{appendix:kinetic}
Before diagonalizing the mass matrix, we must diagonalize the kinetic mixing between the $U(1)$ and $U(1)^\prime$ gauge bosons. This can be done through the following $GL(3,\mathbb{R})$ transformation:
\begin{equation} \label{eq:V}
    \begin{bmatrix}
    W^3_\mu \\
    B_\mu \\
    Z^\prime_\mu
    \end{bmatrix} =
    \mathbf{V} \begin{bmatrix}
    \hat{W}^3_\mu \\
    \hat{B}_\mu \\
    \hat{Z}^\prime_\mu
    \end{bmatrix},
    \hspace{0.3cm}
    \mathbf{V} \equiv \begin{bmatrix}
    1 & 0 & 0 \\
    0 & 1 & -\tan\chi \\
    0 & 0 & 1 / \cos\chi
    \end{bmatrix}
\end{equation}
where the hats indicate fields with canonical kinetic mixing terms. We redefine fields in an analogous way to the (neutral) electroweak sector of the SM:
\begin{equation}
    \begin{bmatrix}
    A_\mu \\
    Z_\mu \\
    Z'_\mu
    \end{bmatrix} =
    \mathbf{W}
    \begin{bmatrix}
    W^3_\mu \\
    B_\mu \\
    Z'_\mu 
    \end{bmatrix}, \hspace{0.4cm}
    \begin{bmatrix}
    \hat{A}_\mu \\
    \hat{Z}_\mu \\
    \hat{Z}'_\mu
    \end{bmatrix} =
    \mathbf{W}
    \begin{bmatrix}
    \hat{W}^3_\mu \\
    \hat{B}_\mu \\
    \hat{Z}'_\mu 
    \end{bmatrix},
\end{equation}
\begin{equation*}
    \mathbf{W} \equiv \begin{bmatrix}
    s_W & c_W & 0 \\
    c_W & -s_W & 0 \\
    0 & 0 & 1
    \end{bmatrix}.
\end{equation*}
Equation \eqref{eq:V} may now be rewritten with the redefined fields:
\begin{equation}
    \begin{split}
    \begin{bmatrix}
        A_\mu \\
        Z_\mu \\
        Z'_\mu
    \end{bmatrix} &= 
    \mathbf{W}
    \mathbf{V}
    \mathbf{W}^{-1}
    \begin{bmatrix}
        \hat{A}_\mu \\
        \hat{Z}_\mu \\
        \hat{Z}'_\mu 
    \end{bmatrix} \\
    &=
    \begin{bmatrix}
    1 & 0 & -c_W \tan\chi \\
    0 & 1 & s_W \tan\chi \\
    0 & 0 & 1/\cos\chi
    \end{bmatrix}
    \begin{bmatrix}
        \hat{A}_\mu \\
        \hat{Z}_\mu \\
        \hat{Z}'_\mu 
    \end{bmatrix}. \label{eq:kinetic_basis}
    \end{split}
\end{equation}

\subsection{Mass Mixing} \label{appendix:mass}
$\mathcal{L}_{\text{mass}}$ admits a $Z_\mu-Z^\prime_\mu$ mass matrix of
\begin{equation*}
    \mathbf{m}^2 =
    \begin{bmatrix}
    m_Z^2 & \Delta^2 \\
    \Delta^2 & m_{Z^\prime}^2
    \end{bmatrix},
\end{equation*}
\begin{align}
    m_Z^2 &= \frac{1}{4} g_Z^2 (|H^0_u|^2 + |H^0_d|^2) \\
    m_{Z^\prime}^2 &= g_{Z^\prime}^2 (Q_{H_u}^{\prime 2} |H^0_u|^2 + Q_{H_d}^{\prime 2} |H^0_d|^2 + Q_S^{\prime 2} |S|^2) \\
    \Delta^2 &= \frac{1}{2} g_Z g_{Z^\prime} (Q'_{H_u} |H^0_u|^2 - Q'_{H_d} |H^0_d|^2)
\end{align}
where $g_Z^2=g_1^2 + g_2^2$ and $|H^0_u|,|H^0_d|,|S|$ are the VEVs of the electrically neutral components of the scalar fields, which we parameterize as $|H^0_u| = v \sin\beta$, $|H^0_d| = v \cos\beta$ with $v=246$ GeV. Anomaly cancellation requires $Q'_{H_u}+Q'_{H_d}+Q'_S=0$.

In the new basis ($\hat{A}_\mu,\hat{Z}_\mu,\hat{Z}'_\mu$) of canonical kinetic terms, the mass matrix $\mathbf{m}^2$ is transformed according to the ($Z_\mu,Z'_\mu$) subspace in Eq. \eqref{eq:kinetic_basis}. The transformation from this subspace to the ($\hat{Z}_\mu,\hat{Z}'_\mu$) subspace is given by
\begin{equation}
    \mathbf{R} =
    \begin{bmatrix}
        1 & s_W \tan\chi \\
        0 & 1/\cos\chi
    \end{bmatrix}
\end{equation}
so that in this basis, $\mathbf{m}^2$ becomes
\begin{equation}
    \begin{split}
    -\mathcal{L}_{\text{mass}}
    &= \begin{bmatrix}
    Z^\mu & Z^{\prime \mu}
    \end{bmatrix}
    \mathbf{m}^2
    \begin{bmatrix}
    Z_\mu \\
    Z^\prime_\mu
    \end{bmatrix} \\
    &=
    \begin{bmatrix}
    \hat{Z}^\mu & \hat{Z}^{\prime \mu}
    \end{bmatrix}
   \mathbf{R}^T
    \mathbf{m}^2
    \mathbf{R}
    \begin{bmatrix}
    \hat{Z}_\mu \\
    \hat{Z}^\prime_\mu
    \end{bmatrix} \\
    &\equiv \begin{bmatrix}
    \hat{Z}^\mu & \hat{Z}^{\prime \mu}
    \end{bmatrix}
    \mathbf{M}^2
    \begin{bmatrix}
    \hat{Z}_\mu \\
    \hat{Z}^\prime_\mu
    \end{bmatrix}
    \end{split}
\end{equation}
where the elements of $\mathbf{M}^2$ are
\begin{align*}
    M^2_{11} &= m_Z^2 \\
    M^2_{12} &= M^2_{21} = \Delta^2/\cos\chi + m_Z^2 s_W \tan\chi \\
    M^2_{22} &= m_{Z^{\prime}}^2/\cos^2\chi + m_Z^2 s_W^2 \tan^2\chi \\
    & \hspace{0.4cm} + 2 \Delta^2 s_W \tan\chi/\cos\chi.
\end{align*}
This new mass matrix $\mathbf{M}^2$ may be diagonalized by an orthogonal matrix
\begin{equation} \label{eq:mass basis}
    \mathbf{O} = \begin{bmatrix}
    \cos\xi & -\sin\xi \\
    \sin\xi & \cos\xi
    \end{bmatrix}
\end{equation}
and the $Z-Z^\prime$ mixing angle $\xi$ is given by
\begin{equation}
    \tan{(2\xi)} = \frac{2 M^2_{12}}{M^2_{11} - M^2_{22}}
\end{equation}
or, in terms of the eigenvalues $M_Z^2$ and $M_{Z^\prime}^2$ of the matrix $\mathbf{M}^2$,
\begin{equation} \label{eq:mixing}
    \sin{(2\xi)} = \frac{2 M^2_{12}}{M^2_Z - M^2_{Z^\prime}}, \hspace{0.4cm}
    \cos{(2\xi)} = \frac{M^2_{11} - M^2_{22}}{M^2_Z - M^2_{Z^\prime}}.
\end{equation}

The nonzero entries of the diagonal mass matrix $\mathbf{O}^T \mathbf{M}^2 \mathbf{O}$ are the tree-level squared masses of the observed $Z$ boson and a new $Z'$,
\begin{align*}
    M_Z^2 &= \frac{1}{2} \left( M^2_{11} + M^2_{22} - \sqrt{(M^2_{11} - M^2_{22})^2 + 4M^2_{12}} \right) \label{eq:MZ} \\
    M_{Z^\prime}^2 &= \frac{1}{2} \left( M^2_{11} + M^2_{22} + \sqrt{(M^2_{11} - M^2_{22})^2 + 4M^2_{12}} \right).
\end{align*}

\subsection{Mass Eigenstates} \label{appendix:eigenstates}

After symmetry breaking,  $\mathcal{L}_{\text{current}}$ becomes
\begin{equation*}
    \mathcal{L}_{\text{current}} = -J^\mu_{\text{em}} A_\mu - J^\mu_Z Z_\mu - J^{\prime \mu} Z^\prime_\mu
\end{equation*}
where these new currents are given by
\begin{align}
    J^\mu_{\text{em}} &= g_1 c_W  \sum_i q_i \Bar{f}_i \gamma^\mu f_i \label{eq:currents_em} \\
    J^\mu_Z &= \frac{g_2}{c_W} \sum_i \Bar{f}_i \gamma^\mu [Q_{ZiL} P_L + Q_{ZiR} P_R] f_i \label{eq:currents_Z}
\end{align}
with
\begin{equation} \label{eq:Zcharges}
    Q_{Zi(L,R)} = T^3_{i(L,R)} - q_i s_W^2
\end{equation}
where $A_\mu=s_W W^3_\mu + c_W B_\mu$ and $q_i \equiv T^3_{iL} + Y_{iL} = T^3_{iR} + Y_{iR}$ is the electromagnetic charge of fermion $f_i$ divided by the positron's charge.

Section~\ref{appendix:kinetic} transforms $\mathcal{L}$ into a basis of fields with canonical kinetic mixing terms through the matrix $\mathbf{W} \mathbf{V} \mathbf{W}^{-1}$. It is then illustrated in Sec.~\ref{appendix:mass} how to transform $\mathcal{L}$ into a basis with a diagonal mass matrix using $\mathbf{O}$. Here, we combine both of these transformations, allowing us to find the mass eigenstate basis:
\begin{equation} \label{eq:mass eigenstates}
    \begin{split}
        -\mathcal{L}_{\text{current}} &= \begin{bmatrix}
        J^\mu_{\text{em}} & J^\mu_Z & J^{\prime \mu}
        \end{bmatrix}
        \begin{bmatrix}
        A_\mu \\
        Z_\mu \\
        Z'_\mu
        \end{bmatrix} \\
        &= \begin{bmatrix}
        J^\mu_{\text{em}} & J^\mu_Z & J^{\prime \mu}
        \end{bmatrix}
        \mathbf{W} \mathbf{V} \mathbf{W}^{-1}
        \begin{bmatrix}
        1 & 0 \\
        0 & \mathbf{O}
        \end{bmatrix}
        \begin{bmatrix}
        \hat{A}_\mu \\
        Z_{1 \mu} \\
        Z_{2 \mu}
        \end{bmatrix} \\
        & \equiv J^\mu_{\text{em}} \hat{A}_\mu + J^\mu_1 Z_{1 \mu} + J^\mu_2 Z_{2 \mu}.
    \end{split}
\end{equation}
We identify the $Z_{1 \mu}$ eigenstate with the observed $Z$ boson and the $Z_{2 \mu}$ eigenstate with a hypothetical $Z'$ boson. The currents coupled to these mass eigenstates are found through Eqs.~\eqref{eq:currents_em}, \eqref{eq:currents_Z} as
\begin{equation}
    J^\mu_n = \sum_i \Bar{f}_i \gamma^\mu [Q_{niL} P_L + Q_{niR} P_R] f_i
\end{equation}
with
\begin{equation} \label{eq:Z1charge}
\begin{split}
    Q_{1i(L,R)} =& g_1 c_W [-c_W \sin\xi \tan\chi] q_i \\
    & + \frac{g_2}{c_W} [\cos\xi + s_W \sin\xi \tan\chi] Q_{Zi(L,R)}  \\
        & + g_{Z^\prime} [\sin\xi / \cos\chi] Q^\prime_{i(L,R)}
\end{split}
\end{equation}
\begin{equation} \label{eq:Z2charges}
\begin{split}
Q_{2i(L,R)} =& g_1 c_W [-c_W \cos\xi \tan\chi] q_i \\
    & + \frac{g_2}{c_W} [-\sin\xi + s_W \cos\xi \tan\chi] Q_{Zi(L,R)}  \\
        & + g_{Z^\prime} [\cos\xi / \cos\chi] Q^\prime_{i(L,R)}.
\end{split}
\end{equation}
The full transformation matrix of Eq.~\eqref{eq:mass eigenstates} is
\begin{equation*}
    \mathbf{W} \mathbf{V} \mathbf{W}^{-1}
        \begin{bmatrix}
        1 & 0 \\
        0 & \mathbf{O}
        \end{bmatrix}=
    \begin{bmatrix}
        1 + \Delta_A & \Delta_{AZ} & \Delta_{AZ'} \\
        \Delta_{ZA} & 1 + \Delta_{Z} & \Delta_{ZZ'} \\
        \Delta_{Z'A} & \Delta_{Z'Z} & 1 + \Delta_{Z'}
    \end{bmatrix}
\end{equation*}
\begin{equation*} \label{eq:matrix}
    =
    \begin{bmatrix}
        1 & -c_W \sin\xi \tan\chi  & -c_W \cos\xi \tan\chi \\
        0 & \cos\xi + s_W \sin\xi \tan\chi & -\sin\xi + s_W \cos\xi \tan\chi \\
        0 & \sin\xi / \cos\chi & \cos\xi / \cos\chi
    \end{bmatrix}
\end{equation*}
and, following Ref.~\cite{Holdom91}, the oblique parameters are given by
\begin{multline}
    \alpha S = 4 c_W s_W [ -(s_W^2 - c_W^2) \Delta_{AZ} \\
    - 2c_W s_W \Delta_{A} + 2c_W s_W \Delta_{Z} ] \label{eq:S}
\end{multline}
\begin{align}
    \alpha U &= -8s_W^2 [-c_W s_W \Delta_{AZ} + s_W^2 \Delta_{A} + c_W^2 \Delta_{Z'}] \label{eq:U}\\
    \alpha T &= 2( \Delta_Z - \Tilde{\Delta}_Z ) \label{eq:T}
\end{align}
where $\tilde{\Delta}_Z$ is the $Z$ boson's fractional mass shift from its SM value, derived in Sec.~\ref{Sec:mW}.
\vspace{11.5mm}

\bibliographystyle{apsrev4-1.bst}
\bibliography{mW.bib}
\end{document}